\documentclass[twoside]{article}
\usepackage{fleqn,espcrc2}

\newcommand{\beq}{\begin{equation}}
\newcommand{\eeq}{\end{equation}}
\newcommand{\beqn}{\begin{eqnarray}}
\newcommand{\eeqn}{\end{eqnarray}}
\newcommand{\bea}[1]{\begin{array}}
\newcommand{\eea}{\end{array}\ee}

\newcommand{\NPPS}[3]{{\it Nucl. Phys. Proc. Suppl. }{\bf #1} (#2) #3}

\newcommand{\PTPS}[3]{{\it Prog. Theor. Phys. Suppl. }{\bf #1} (#2) #3}

\title{From confining fields back to   power corrections}
\author{V.I. Zakharov\address{
	Max-Planck Institut f\"ur Physik, \\
	F\"ohringer Ring 6, 80805 M\"unchen, Germany.}}
\begin{document}
\begin{abstract}
{We overview the issue of the power corrections to the parton model,
mostly within the context of QCD sum rules. There are a few sources
of the power corrections responsible for various qualitative effects. 
The basic idea that the power corrections are related to confining
fields seems to be realized in case of $1/Q^2$ corrections which
are dual to high orders of perturbation theory. The evidence comes 
from lattice studies of the confinement which allowed to identify
the confining field configurations as monopoles and vortices. 
We comment also on the role of the condensate of dimension two, 
$<(A_{\mu}^a)^2>_{min}$.}
\end{abstract}

\maketitle

\section{INTRODUCTION}

The issue of the power corrections within the context of the QCD sum rules \cite{svz}
has been subject of many reviews, see, e.g., \cite{narison,gpz,shuryak}.
Many things have become common knowledge and there is no need to review
them further. Moreover, we actually assume that 
the reader is familiar with sum rules, at least, in general terms.
One of the main questions, however, has not been yet fully resolved.
Namely, the power corrections were introduced first \cite{svz}
to parametrized the effect of confining fields. 
Nowadays, however, 
this issue is overshadowed by technical details of application of the sum rules.
It might be still useful to evaluate the status of this issue. 

Actually, it is at this point that the situation now is radically different from
the early years of the sum rules. Namely, there was little hope that the confining fields 
could be identified explicitly since confinement is a property of strong-coupling theory.
Thus, the sum rules were intended to at least systematically parameterize  effects
of confinement at short distances. Nowadays, lattice studies seem to reach the point
when the knowledge of confining field configurations is quite precise. 
The confining field configurations are the lattice monopoles and vortices, 
for a recent review see, e.g.,
\cite{greensite}. Thus, one can try to project these fields back to short distances and
clarify what kind of power corrections they generate.
We will argue that
it is the so called quadratic corrections \cite{chetyrkin,akhoury,gpz}
which seem to be directly related to the confining fields \cite{vz}.

The issue of the quadratic corrections is somewhat controversial.
On the scientific side, there
is a well defined reason for this. In terms of the Operator Product Expansion (OPE)
the quadratic corrections are a part of the coefficient functions in
front of the unit operator and are not related to a matrix element
of a non-trivial operator. In other words, the quadratic corrections are
actually dual to high orders of perturbation theory. This is a unique feature of these
confinement-related power corrections and the common culture of working with
the sum rules does not help much to appreciate this special feature of
 the quadratic corrections.

On the other hand, this manifestation of duality in pure Yang-Mills theories
might help in understanding dual formulations of Yang-Mills
theories which is commonly believed to be a string theory, see, in particular,
\cite{maldacena}.

\section{SUM RULES}
\subsection{Classical channels}

In somewhat simplistic form, the sum rules read:
\beqn\label{sumrules}\nonumber
\int ds\exp(-s/M^2)R_j(s)~\approx~(parton~ model)\cdot\\
\cdot\big(1+a_j\alpha_s(M^2)+b_j<G^2>/M^4\big)
\eeqn
where $M^2$ is a parameter, $M^2\gg \Lambda_{QCD}^2$, the coefficients $a_j,b_j$ 
are calculable perturbatively,
$R_j$ is proportional 
to cross section induced by a hadronic current $j$
(if, say, $j$ is the electromagnetic current then $R_j$ is proportional to
the total cross section of $e^{+}e^{-}$-annihilation), $\alpha_s(M^2)$
is the running coupling, $<G^2>$ is the gluon condensate. There are actually
higher power corrections, say, proportional to $M^{-6}$ which we omit here.

It might worth emphasizing that the sum rules (\ref{sumrules})
are far from being `natural'. 
Indeed, we cut short the perturbative series by keeping only the first
perturbative correction, $a_j\alpha_s(M^2)$ and, on the other hand,
keep the power corrections. However, for very small $\alpha_s$ any
power of $\alpha_s$ is much larger than any power correction since
$M^{-2}\sim \exp(-b_0/\alpha_s(M^2))$. 

The idea behind this breaking of the perturbative hierarchy was that
the pure perturbative corrections are not signalling confinement.
And the physics of confinement is encoded in the power corrections
which become crucial at small, but not very small $\alpha_s$ 
and are synchronized with deviations from the parton model in the
measurable cross section $R_j$.

There is no a priori justification for such an assumption and one relies entirely
on the phenomenological success to 
pursue the sum rules in the form (\ref{sumrules}).
In particular the sum rules in the $\rho$- and $J/\Psi$-channels allowed
to `explain' the approximate equality of the mass splittings:
\beq\label{equi}
m_{\Psi^{'}}-m_{J/\Psi}~\approx~m_{\rho^{'}}-m_{\rho}~~.
\end{equation}
Although Eq (\ref{equi}) looks trivial it is actually difficult to understand
theoretically since relativistically we should compare the 
 masses squared, and these are very different in different channels. 
Sum rules based on the asymptotic freedom alone \cite{six} fail to explain (\ref{equi})
completely. Explaining (\ref{equi}) was the first qualitative success of the sum rules
(\ref{sumrules}).

\subsection{Hierarchy of scales}

Sum rules (\ref{sumrules}) have numerous successful applications.
There are, however, remarkable failures as well \cite{scales,shuryak}.
Namely, the most straightforward prediction of the sum rules is the value
of $M^2_{crit}$ at which violations of the parton model become
sizable. The numbers vary from channel to channel
because the coefficients like $b_j$ (see (\ref{sumrules})) depend on
the channel.

A careful (independent of the sum rules) phenomenological 
analysis \cite{scales} allowed
to establish a kind of hierarchy of scales:
\beqn\label{hierarchy}
\big(M^2_{crit}\big)_{\pi}~\approx~4\cdot 
\big(M^2_{crit}\big)_{\rho}\\\nonumber
\big(M^2_{crit}\big)_{glueball~0^{+}}~\approx~4\big(M^2_{crit}\big)_{\pi}
\eeqn
The standard sum rules (\ref{sumrules}) fail completely to reproduce this
hierarchy.

\subsection{Direct instantons}

There is one quite obvious reason to modify the sum rules (\ref{sumrules}).
Indeed, none of the terms kept in (\ref{sumrules}) would signal 
special role of the $\eta^{'}$-channel. It is known since long
\cite{thooft} that one has to include instantons to explain shifting
of the $\eta^{'}$-meson mass compared to other pseudoscalar mesons.

The instanton contribution at large $M^2$ starts as a highly suppressed power
correction \cite{nsvz} to the right hand side of Eq (\ref{sumrules}):
\beq
\delta_{instantons}~\sim~\big({\Lambda_{QCD}\over M}\big)^9~~.
\eeq
However, at values of $M^2$ which become sensitive to the
resonance structure in the left-hand side of (\ref{sumrules}) this
correction should overcome other sources of the power corrections,
at least in case of some channels, like the $\eta^{'}$-channel.

Detailed analysis of the instanton contribution relies actually on 
elaborated models of the vacuum, like instanton-liquid model, for review
see, e.g., \cite{shuryak}. There is overwhelming evidence in favor of
the instantons in some channels.

\section{QUADRATIC CORRECTIONS}
\subsection{Short-distance gluon mass}

Even upon inclusion of the instantons there remain phenomenological 
problems with the sum rules. The example easiest to
understand is provided by the  heavy quark static potential at short distances. Strictly
speaking, the potential cannot be treated within the 
(OPE). However, quite a straightforward extension of the OPE predicts
that the correction due to the soft non-perturbative fields is of
order \cite{balitskii}
\beq\label{balitskii}
\delta V_{Q\bar{Q}}(r)~\sim~\Lambda_{QCD}^3\cdot r^2
\eeq
where $r$ is the distance between the quarks and $r$ is small, $r\cdot\Lambda_{QCD}\ll 1$.

However, the lattice data are fitted by the so called Cornell potential,
\beq\label{unusual}
V_{Q\bar{Q}}(r)~\approx~-{const\cdot\alpha_s\over r}+\sigma\cdot r~~,
\eeq
at all distances available. In particular, if one trusts this fit at short distances
there is a linear in $r$ correction to the Coulomb-like
potential, $\delta V\sim \sigma\cdot r$ \cite{akhoury}.

The argumentation just presented is actually not so strong. The problem--quite usual
one with the power corrections-- is that one should confine oneself to the distances
where the power correction to the Coulomb-like potential
is small. But then  it is difficult to extract the
hypothetical piece $ \delta V\sim \sigma\cdot r$ on the background of the 
perturbative corrections, see, in particular, \cite{bali}.

The only way out, to my mind, is to analyze along the same lines
a few channels or find qualitative effects. In case of the quadratic corrections,
unfortunately, there is no rigorous way to relate corrections in various channels.
Indeed, so far we have not yet even identified the source of such corrections.

An attempt to break the deadlock was made in ref. \cite{chetyrkin}
by introducing a model to relate quadratic correction in various channels.
The starting observation is that to reproduce the potential (\ref{unusual}) 
at short distances one can modify the gluon propagator at short distances
by introducing the gluon mass:
\beq\label{mass}
1/Q^2~\to 1/Q^2+m_g^2/Q^4~~,
\eeq
where $Q$ is the Euclidean momentum carried by gluon and $Q^2\gg \Lambda_{QCD}^2$.
Ones replacement (\ref{mass}) is postulated, the quadratic correction is calculable
in other channels as well.

It turns out that the model (\ref{mass}),
using as an input the string tension $\sigma$ immediately reproduces
both the overall scale and 
the hierarchy of the anomalous contributions
(\ref{hierarchy}) \cite{chetyrkin}. There are a few other non-trivial checks
of the model. Note also that $m_g^2$ in Eq (\ref{mass}) is positive
which implies that the short-distance gluon mass is in fact tachyonic.
Later the prediction of the model were scrutinized further \cite{narison1}
and so far the model works very well, for a review see the talk by
S. Narison at this Conference \cite{narison2}

\subsection{Ultraviolet renormalon}

In view of the phenomenological success of the model (\ref{mass}) it is 
useful to clarify the theoretical status of the model.
Historically, the power corrections were introduced first
within the context of divergences of perturbative series, for review see,
e.g., \cite{beneke1}.

Consider, for example, polarization operator $\Pi_j(Q^2)$
 associated with a hadronic current $j$. Perturbatively:
\beqn\nonumber
\{\Pi_j(Q^2)\}_{pert}=(parton~model)\cdot\\
~~~~~~~~~~~~~~~~~\cdot\big(1+\Sigma_{n=1}^{\infty} a_n(\alpha_s(Q^2))^n\big)~.
\eeqn
The sum is however formal because of the factorial growth of the
expansion coefficients in large orders:
\beqn\label{first}
 (a_n)_{UV}~\sim~(-1)^nn!b_0^{n}\\ \label{second}
(a_n)_{IR}~\sim~n!2^{-n}b_0^{n}~~,
\eeqn
where $b_0$ is the first coefficient in the $\beta$-function.
Moreover, the series (\ref{first}) is associated with integration
over virtual momenta $k^2\gg Q^2$ while the series (\ref{second}) is due to
$k^2\ll Q^2$. Assume that the perturbative series is asymptotic and introduce
corresponding power corrections as estimate of uncertainties associated
with the divergence (\ref{first}) and (\ref{second}). It is straightforward to see
that the corrections are of order $\Lambda^2_{QCD}/Q^2$ and $\Lambda_{QCD}^4/Q^4$,
respectively.

Now, the central point is that (\ref{first}) in fact does not introduce any
uncertainty. The simplest way to argue so is to recall that (\ref{first}) is Borel summable.
A more elaborated argumentation \cite{beneke} exploits the fact that at large
virtual momenta calculations in QCD are reliable because of the asymptotic freedom.

\subsection{Short and long perturbative series}

Let us emphasize again that originally sum rules contain short
perturbative series, see Eq. (\ref{sumrules}). The $1/Q^2$ correction was introduced within this
context. In fact, the replacement (\ref{mass}) is to be used in the Born approximation and
is not suited, say, for the renormalization program. 
 If, however, one calculates many terms in the series, no $1/Q^2$ term can be added
since the ultraviolet renormalon is calculable.

From the point of view of applications, the crucial question is, how many
orders one should calculate explicitly to avoid adding $1/Q^2$ correction. 
Unfortunately, there exists no theoretical framework
to answer this question and
the answer may vary from channel to channel.
Numerically, the gluon condensate is studied 
by far much better than any other observable \cite{rakow}.
Perturbatively:
\beqn\nonumber
\langle 0|{\beta(\alpha_s)\over \alpha_s}(G_{\mu\nu}^a)^2|0\rangle_{pert}~=~\\
~~~~~~~~~~~~~~~~{const\over a^4}\big[1+\Sigma a_n\alpha_s^n(1/a)\big]~~,
\eeqn
where $1/a\equiv\Lambda_{UV}$ is the ultraviolet cut off ($a$ is the lattice spacing)
and in the terminology we used above, $Q^2\to 1/a^2$.
First 10 terms of the perturbative expansion were calculated
explicitly and there is no sign yet that the infrared renormalon sets in. In other words, 
the $1/Q^2$ correction is still to be added by hand to describe the data.
Extrapolation of the coefficients obtained to higher $n$ indicates, 
however, that finally the $1/Q^2$ is absorbed into the perturbative series,
as is expected theoretically.

\subsection{Comments on literature}

For many years, dominance of soft-field corrections to the heavy quark-potential
at short distances was considered obvious, see, e.g. \cite{nora} \footnote{
In case of bound states of heavy quarks, the effects of retardation \cite{peskin}
reduce the correction (\ref{balitskii}) by an extra power of
$(\Lambda_{QCD}\cdot r)$  \cite{voloshin}.}.
On this background, introduction of a linear term in the
potential at {\it short distances} \cite{chetyrkin,akhoury,gpz,gpz1} looked heretical.

Nowadays, it seems to become a common point that the soft-field contribution 
to the heavy-quark potential (\ref{balitskii}) is negligible
and effective linear term is there, see, e.g., \cite{pineda} . Moreover, a careful
analysis of Ref. \cite{bali1} also supports the guess (\ref{mass}) that the short distance
$1/Q^2$ correction is significant specifically 
in one-gluon channel. Once a combination of channels is considered which is not
contributed by a one-gluon exchange the soft-field contribution
might show up \footnote{After finding this confirmation
of \cite{gpz,chetyrkin} the authors 
of Ref. \cite{bali1} summarize their results as being
in contradiction with these papers. This is a clear misstatement.
The author is thankful to G. Bali and A. Pineda for correspondence
on this point.}. 
In this sense, the phenomenological support for the models \cite{chetyrkin,gpz,akhoury}
has been strengthened recently.

However, the recent trend \cite{pineda} is to work with long perturbative series
which presumably absorb the $1/Q^2$ corrections. For this reason, direct comparison between \cite{pineda}
and models \cite{chetyrkin,gpz} formulated within the context of short
perturbative series is not straightforward. This issue of duality between high
orders of perturbation theory and the quadratic corrections is indeed crucial
to compare various approaches.

It is worth mentioning that the recent papers \cite{pineda}
concentrate exclusively on the heavy-quark potential. No actual
`long enough' perturbative series is known explicitly in this case.
To substitute for this lack of knowledge,
extrapolations utilizing infrared renormalons are common.
A priori, such extrapolations could be correct or wrong
\footnote{To our knowledge, there is no single example so far 
of an explicit
calculation of a perturbative series which would exhibit set in of the
renormalon-related asymptotic of the coefficients. The longest series known
\cite{rakow} is no exception in this sense. For the most recent review of
the status of renormalons and their possible generalization see \cite{kataev}.}.

The actual check of the models \cite{pineda}, to our mind, would be 
to apply these models in other channels, see, e.g., (\ref{hierarchy}).
It goes without saying that comparison of extrapolations with the
explicit long series available in case of the gluon condensate \cite{rakow}
would be most desirable.

\section{CONFINING FIELDS}
\subsection{Monopoles, vortices}

Painstaking search for confining fields in lattice Yang-Mills theories took
many years but was crowned with success. Namely the confining fields can be viewed either
as lattice monopoles, for review see \cite{monopoles} or vortices, for review see \cite{greensite}.
Moreover, the difference between the two languages is superficial,
see, e.g.,   \cite{kovalenko}.
Namely, the confining field configurations are 2d surfaces (vortices)
populated with particles (monopoles).

Usually monopoles and vortices are treated as effective infrared degrees of freedom
\cite{monopoles,greensite}. However, they have an ultraviolet facet as well \cite{anatomy,kovalenko}.
In particular, for the vortices one finds:
\beq\label{facet}
A~\approx~24(fm)^{-2}V_4~\cite{greensite},~S~\approx~0.5{A\over a^2}~\cite{kovalenko},
\eeq
where $A$ is the total area of the vortices, $S$ is non-Abelian action associated
with the vortices, $V_4$ is the total volume. Note that the action diverges in
the continuum limit of the vanishing lattice spacing, $a\to 0$.

\subsection{Back to the quadratic corrections}

Both monopoles and vortices are extended geometrical objects. 
However, if they are so to say projected onto
a (quasi)local variable their contribution corresponds to the quadratic correction
considered above \cite{vz}. This can be explicitly demonstrated in at least two cases.
Namely, combining the two observations cited in (\ref{facet}) 
we get for the contribution of the vortices into the gluon condensate:
\beq
\langle 0|\alpha_s(G_{\mu\nu}^a)^2|0\rangle_{vortices}~\sim~{const\over a^4}(\Lambda_{QCD}\cdot a)^2~~,
\eeq
which is nothing else but a quadratic correction to the perturbative result.

Moreover, monopoles and vortices generate linear piece in
the heavy quark potential at short distances:
\beq\label{gpz}
\lim_{r\to 0}{\delta V_{Q\bar{Q}}(r)}~\sim~\sigma\cdot r
\eeq
where $\sigma$ numerically is very close to the (large-distance) string tension.
Theoretical reasoning for validity of (\ref{gpz})
was given first in Ref. \cite{gpz1} where references to the data can also
be found.

\subsection{Duality}

It is worth emphasizing again that the contribution of the vortices to 
local observables should not be added to perturbation theory but is
dual to high orders of perturbation theory. In case of the gluon
condensate this theoretical prediction has been already verified.
Indeed the difference between the full value of the gluon 
condensate and its perturbative value is known explicitly \cite{rakow}
to be of order:
\beqn\label{difference}\nonumber
\langle 0|\alpha_s(G_{\mu\nu}^a)^2|0\rangle_{total}-
\langle 0|\alpha_s(G_{\mu\nu}^a)^2|0\rangle_{pert}\\
~~~~~~~~~~~~~``\sim\Lambda_{QCD}^4~.
\eeqn
Note that (\ref{difference}) is valid only if more than ten orders
of perturbation theory are subtracted while for shorter perturbative series
it is the quadratic correction that dominates the difference (\ref{difference}).

This duality in evaluating a particular
observable might indicate existence of a dual formulation of the Yang-Mills 
theory itself \cite{vizz}. Indeed, both vortices and, say, instantons are field configurations
defined in terms of the original Yang-Mills fields. Nevertheless instantons are
responsible for terms of order $\Lambda_{QCD}^4$ in (\ref{difference}) and they
should be added to the perturbative series. While the contribution of the vortices
is dual to perturbation theory. The difference can be understood if one assumes that
the vortices become fundamental variables of a dual formulation of 
the same Yang-Mills theories. 

Indeed, then we should have two alternative representations for the
same heavy-quark potential, in terms of the fields and in terms of
the strings. In particular the potential at large distances is trivial in terms of strings:
\beq
\lim_{r\to \infty}V_{Q\bar{Q}}\sim(strings~contribution)
~~\eeq
while no close expression is known in terms of fields, i.e. perturbatively.
Similarly, at short distances the potential is trivial in terms of the fields, i.e. of the gluon exchange:
\beq
\lim_{r\to 0}V_{Q\bar{Q}}~\sim~(fields~contribution)~~,
~~\eeq
with no close expression in terms of the strings being expected.
At intermediate distances, where the linear term is visible
the potential is calculable in two alternative ways, either
in terms of strings or perturbative series:
\beqn
\delta V_{Q\bar{Q}}(r)~\sim~\sigma\cdot r
~\sim~(fields~ contribution)\\
~~or~\sim~(strings~contribution)~.
\eeqn 
Adding the two
is not allowed.

\section{DIMENSION-TWO CONDENSATE}
\subsection{OPE and condensate of dimension two}

The power corrections in the sum rules (\ref{sumrules}) start with $1/Q^4$
term, with no $1/Q^2$ contribution. The reason is that the first gauge-invariant operator,
that is $(G_{\mu\nu}^a)^2$, has dimension $d=4$.

As a matter of fact, dimension two condensate can be introduced \cite{a2}
in a gauge invariant fashion.
Indeed, consider first for simplicity magnetostatics and imagine that there exist thin tubes 
carrying magnetic flux. Then the `vacuum expectation value' $\langle 0|{\bf A}^2|0\rangle$
cannot vanish. Moreover, its minimal value is obviously related to the magnetic fluxes and
gauge invariant. Generalizing this logic, one can define gauge invariant condensate of dimension two
as the minimal value of $\langle 0|(A_{\mu}^a)^2|0\rangle$ in Euclidean space-time.  
The meaning of this definition can be scrutinized in great detail \cite{a2}.

Moreover, the value $\langle A^2\rangle$ enters the OPE for gauge {\it variant}
quantities, see Ref. \cite{orsay} and references therein.
However, the condensate $\langle A^2\rangle_{min}$ does not enter
the OPE for gauge invariant quantities. Imagine that one can calculate all terms 
in the OPE for a polarization operator $\Pi_j(Q^2)$ induced by a gauge invariant current
$j$,
\beq\label{ope}
\Pi_j(Q^2)~(parton~model)(1+\Sigma_n (\Lambda_{QCD}/Q)^n)~,
\eeq
where $\Lambda_{QCD}$ to a corresponding power is associated with matrix elements
of gauge invariant operators and, for simplicity, we omitted perturbative series.
The matrix element $\langle (A_{\mu}^a)^2\rangle_{min}$ does not
appear in the r.h.s. of Eq. (\ref{ope}).

Seemingly, this observation is confused sometimes for a proof that the condensate of dimension two
is irrelevant to gauge invariant quantities, say, hadronic masses.
There is no such proof, however. The point is that the expansion (\ref{ope}) is
asymptotic at best. Thus, it is not a closed expression and (\ref{ope}) cannot determine physical
masses in a model independent way, as a matter of principle. From this point of view,
it might be even `good' for $\langle A^2\rangle_{min}$ that it does not enter (\ref{ope}).
Relation of the type, $m^2_{hadron}~\sim~\langle A^2\rangle_{min}$ is still perfectly
possible.

\subsection{$\langle A^2\rangle_{min}$ as  Higgs condensate}

Thus, it is not surprising at all that there appear models
which relate $\langle A^2\rangle_{min}$ to physical observables and 
these models turn to be successful, see, e.g., \cite{pheno}.

Let me mention specifically an idea that $\langle A^2\rangle_{min}$ plays
a role of a Higgs field. The problem is that the lattice monopoles are
defined
usually \cite{monopoles} within an Abelian projection of the original
Yang-Mills fields. In this projection, magnetically-charged field plays the role of the field
describing Cooper pairs in theory of superconductivity. This is not gauge invariant
description however. Moreover, absence of gauge invariant condensate of dimension two
might be thought of as a general objection to the dual-superconductor model of confinement.
It was conjectured in Ref. \cite{higgs} that $\langle A^2\rangle_{min}$ plays
the role of Higgs field in the Landau gauge. The guess is supported by measurements
of variation of $\langle A^2\rangle_{min}$ inside the confining tube,
as function
of the distance to the central axis of the tube. It turns out that the penetration
length determined in terms of this variation is very close numerically
to the penetration length in the Maximal Abelian gauge determined in terms
of the monopole distribution. Confirming the idea that
$\langle A^2\rangle_{min}$ substitutes for the Higgs condensate.

\section{CONCLUSIONS}

We have argued that quadratic corrections to the sum rules are directly related 
to the confining fields. In terms of the OPE, the quadratic correction 
is a part of the coefficient function
in front of the unit operator. Thus, to judge on the value of the correction
one should distinguish models with `short' and `long' perturbative series
kept explicit. In other words, the quadratic correction is dual to high orders
of perturbation theory.
As far as models are concerned, the model with short-distance tachyonic mass
(and short perturbative series) seems to be most successful phenomenologically.

\section{ACKNOWLEDGEMENTS}
We are thankful to S. Narison, L. Stodolsky,  Y. Sumino,
T.~Suzuki for useful discussions.

\end{document}